\documentclass[12pt]{iopart}
\usepackage{graphicx} 
\usepackage{soul}
\usepackage{ulem}
\usepackage{xcolor}
\usepackage{amsfonts} 
\usepackage[makeroom]{cancel}
\usepackage{iopams} 
\usepackage{txfonts}

\newcommand{\ou}{
  \mathrel{
    \vcenter{\offinterlineskip
      \ialign{##\cr$\prec$\cr\noalign{\kern-1.5pt}$\succ$\cr}
    }
  }
}

\begin{document}

\title{Globally hyperbolic evaporating black hole and the information loss issue}

\author{Juan V. O. P\^egas}
\address{Instituto de F\'\i sica Te\'orica, Universidade Estadual Paulista, Rua Dr. Bento Teobaldo Ferraz, 271, 01140-070, S\~ao Paulo, SP, Brazil}
\ead{jv.pegas@unesp.br}

\author{Andr\'e G. S. Landulfo,}
\address{Centro de Ci\^encias Naturais e Humanas, Universidade Federal do ABC, Rua Santa Adélia, 166, 09210-170, Santo Andr`e - SP - Brazil.}
\ead{andre.landulfo@ufabc.edu.br}

\author{George E. A. Matsas\footnote{Corresponding author}}
\address{Instituto de F\'\i sica Te\'orica, Universidade Estadual Paulista, Rua Dr. Bento Teobaldo Ferraz, 271, 01140-070, S\~ao Paulo, SP, Brazil}
\ead{george.matsas@unesp.br}

\author{Daniel A. T. Vanzella}
\address{Instituto de F\'\i sica de S\~ao Carlos, Universidade de S\~ao Paulo, Caixa Postal 369, 13560-970, S\~ao Carlos, SP, Brazil.}
\ead{vanzella@ifsc.usp.br}

\vspace{10pt}

\begin{abstract} 
We discuss the information loss issue for completely evaporating black holes in the context of a globally hyperbolic spacetime that maintains unchanged the entire semiclassical picture except for the ``last evaporation breath,'' which pertains to full quantum gravity. Even though observers outside the black hole cannot access information that enters the horizon, there is no actual loss of information imposed by the spacetime causal structure since it is carried over from one Cauchy surface to the next (provided the local evolution is unitary). 
\end{abstract}

\maketitle

     \section{Introduction}
     \label{introduction}

The loss of information in black holes has attracted much attention in the last 50 years. Generally speaking, there are two large parties today~\cite{M09}. One of them, championed by Wilczek and Maldacena, among others, considers the information loss in black holes and corresponding entropy increase paradoxical. Wilczek et al defend that the total information could be partially~\cite{Parikh00} or totally~\cite{Zhang09} preserved by enforcing energy conservation along the black hole evaporation process, while Maldacena et al~\cite{Almheiri21} try to describe black holes from the point of view of external observers as standard thermodynamic systems (an approach known as the ``central dogma"). Quoting Almheiri et al~\cite{Almheiri21},   
\begin{quotation}
 ``How is a black hole different from a hot piece of coal? This central dogma is saying that, as long as you remain outside, it is not fundamentally different [...] It involves a certain dose of belief because it is not something we can derive directly from the gravity description. We can view it as an unproven assumption about the properties of a full theory of quantum gravity. It is also something that is not accepted by some researchers. In fact, Hawking famously objected to it; [...] In other words, if a black hole is a hole in space where things can get in and get lost, then the central dogma would be false.''   
\end{quotation}
   The other party championed by Hawking~\cite{H76}, Wald~\cite{W94book}, Unruh~\cite{UW95, UW17},  Penrose~\cite{P04book}~(pp. 840-841), among others, composed of skeptics about the ``central dogma,'' sees the information loss simply as a prediction of quantum field theory in curved spacetimes. Quoting Wallace~\cite{W18}, 
\begin{quotation}
``The stripped-down version of their arguments would be: we have a right to expect unitarity, information preservation, and retrodiction only on globally hyperbolic spacetimes; the evaporation spacetime manifestly is not globally hyperbolic; so non-unitary evolution is only to be expected.''  
\end{quotation} 
We belong to the skeptical party. Thus, we will not look for mechanisms to recover the information that crosses the event horizon here. Instead, we elaborate on a globally hyperbolic spacetime describing the complete evaporation of a black hole into Hawking radiation (in the sense that no massive remnant is left over at the end). In this way, ``information loss'' is avoided; not because Hawking radiation would carry information away (after some Page time~\cite{P93,P13}), but because every one of the nonsingular spatial sections used to foliate the spacetime (Cauchy surfaces) — which at ``late times'' necessarily intersects the black hole interior — contains the whole information. In a certain proper sense, the black hole can always be blamed for the missing information in the outside region, despite its complete evaporation.  
 
The paper is organized as follows. In Sec.~\ref{sec:paradox} we revisit the black hole information loss issue. In Sec.~\ref{sec:Hiscock} we review the main features of a spacetime associated with a light-like collapsing spherical shell and the corresponding full evaporation of the black hole into thermal radiation, Figs.~\ref{fig:Hiscock1} and~\ref{fig:Hiscock2}.  In Sec.~\ref{sec:PLMV} we see how the metric must change to replace the final naked singularity with a null singularity, Figs.~\ref{fig:PLMV1} and~\ref{fig:PLMV2}, and discuss the main features of the spacetime. In Sec.~\ref{Page} we discuss the central assumption for the existence of a Page time after which information would be returned to the exterior of a black hole. In Sec.~\ref{Final} we summarize our conclusions. We assume metric signature $(- + + +)$ and natural units $\hbar=G=c=k_B=1$, unless stated otherwise.

     \section{On the information loss in black holes}
     \label{sec:paradox}

\begin{figure}[th]
       \centering
       \includegraphics[width=90mm]{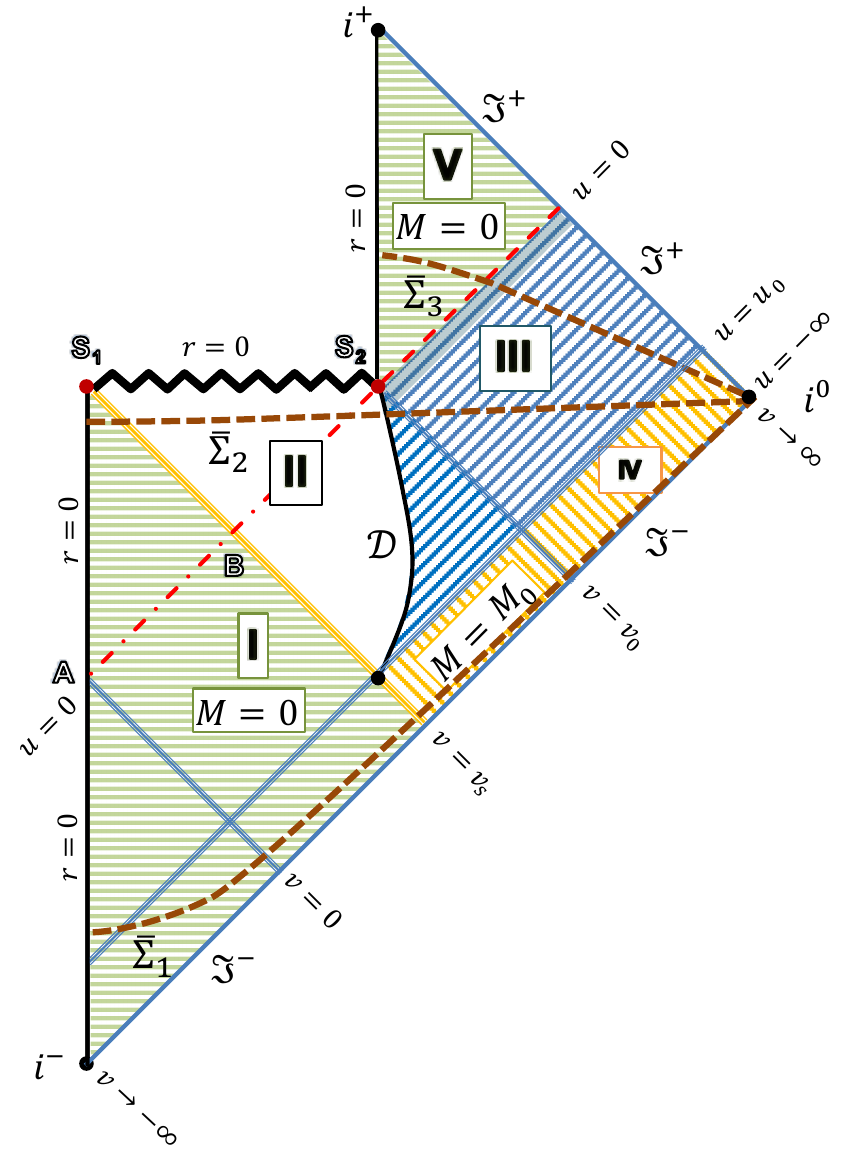}
       \caption{Conformal Carter-Penrose diagram associated with the total collapse of a spherically symmetric light shell ($v=v_s$). This leads to a black hole with mass~$M_0$ that evaporates into Hawking radiation. The spacetime inside the shell, region~I, is flat. Event~$\mathbb{B}$ characterizes the crossing of the ingoing collapsing shell with the outgoing EH.  The spacetime is not globally hyperbolic, and the dashed spacelike hypersurfaces~$\overline{\Sigma}_i$, $i=1,2,3$, are not Cauchy surfaces. The timelike curve~${\cal D}$, ending at the naked singularity~$\mathbb{S}_2$, separates regions~II and~III approximated by metrics~(\ref{ingoing metric}) and~(\ref{outgoing metric}) of ingoing negative-energy and outgoing positive-energy Hawking radiation, respectively.  Region~IV represents a distant region where the spacetime is well approximated by a vacuum Schwarzschild metric. Hawking radiation is depicted by the translucent strip $u \lesssim 0$ at region~III, and is assumed to carry out all black hole energy.  Region~V is a Minkowski spacetime in its own right, which emerges after the complete black hole evaporation. 
       }  
       \label{fig:Hiscock1}
\end{figure}

\begin{figure}[th]
       \centering
       \includegraphics[width=110mm]{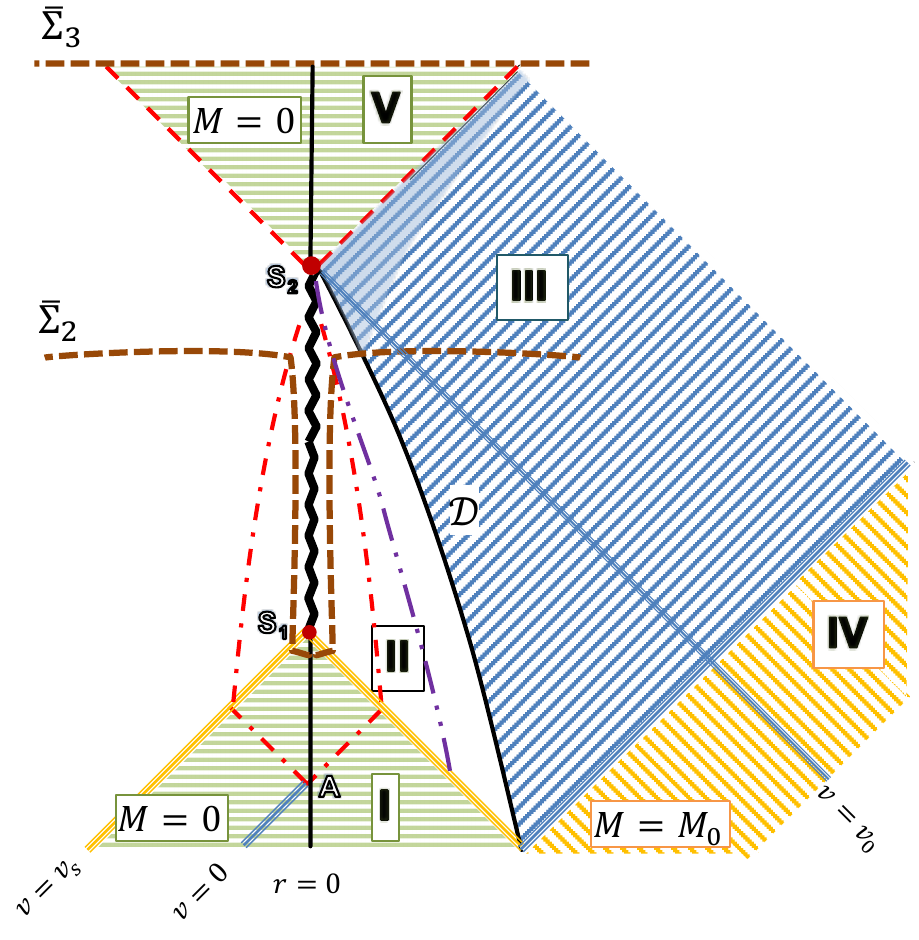}
       \caption{Sketch of the Finkelstein diagram corresponding to the Carter-Penrose one shown in Fig.~\ref{fig:Hiscock1}. The diagram is axially symmetric around the $r=0$ axis, but we have avoided repeating information in order not to overload the figure. One can see that $\overline{\Sigma}_2$ cannot be continuously deformed into $\overline{\Sigma}_3$ without crossing the singularity. There is nothing impeding information on $\overline{\Sigma}_2$ dropping in the singularity instead of reaching $\overline{\Sigma}_3$. We also show, here, the AH contained in region~II depicted with a dashed-double-dotted line.   
       }  
       \label{fig:Hiscock2}
\end{figure}

From a quantum-field theoretic perspective, the loss of information in black holes reflects the fact that physical (Hadamard) states of quantum field theory are in general entangled over space-like separated regions, leading pure states to be perceived as mixed states by observers outside the black hole~\cite{H76, UW17}. The distress that is qualified under the name ``black hole information-loss paradox'' comes from the unusual feature that a pure state prepared in the Minkowski-like asymptotic past would eventually be described as a mixed state in the Minkowski-like asymptotic future after the black hole has completely evaporated. This can be already seen in the familiar context of non-relativistic quantum mechanics, as follows: consider a two-particle entangled state 
\begin{equation}
    |\psi \rangle 
    \equiv 
    \frac{1}{\sqrt{2}} |\uparrow_1 \; \rangle  \otimes | \downarrow_2 \;\rangle 
    +
    \frac{1}{\sqrt{2}} |\downarrow_1 \; \rangle  \otimes | \uparrow_2 \;\rangle
    \label{initial state}
\end{equation}
prepared somewhere in region~I of Figs.~\ref{fig:Hiscock1} and~\ref{fig:Hiscock2} and outside the black hole. Let us, now, unitarily evolve $| \psi \rangle$ such that particle~1 enters the hole while particle~2 proceeds to the asymptotic timelike future~$i^+$. Having no access to the degrees of freedom of particle~1, observers dwelling in region~V would describe the quantum system as a mixed state given by the density matrix 
\begin{equation}
    \hat\rho
    =
    \frac{1}{2} \; |\uparrow_2   \; \rangle \langle \; \uparrow_2| 
    +
    \frac{1}{2} \; |\downarrow_2 \; \rangle \langle \; \downarrow_2|.
    \label{final state}
\end{equation}
As a consequence, there is no way to recover the initial state~(\ref{initial state}) from the final state~(\ref{final state}).

From a quantum-field-theory-in-curved-spacetime perspective, the puzzle vanishes after noticing that the quantum system is not isolated; namely, the spacelike singularity ``opens the system'' by destroying information.  Supporters of the ``central dogma''~\cite{Almheiri21}, however, resist accepting that black holes would behave differently from, say, a piece of coal for observers outside the horizon. Yet, past attempts to harmonize black holes with usual thermodynamical systems for external observers have led to all sorts of undesired features, such as a violation of causality and other basic principles in low curvature regimes (see Ref.~\cite{UW17} for a general overview, and Ref.~\cite{Gan20} for specific issues concerning the firewall conjecture). 

\begin{figure}[htbp]
    \centering
    \includegraphics[scale=0.68]{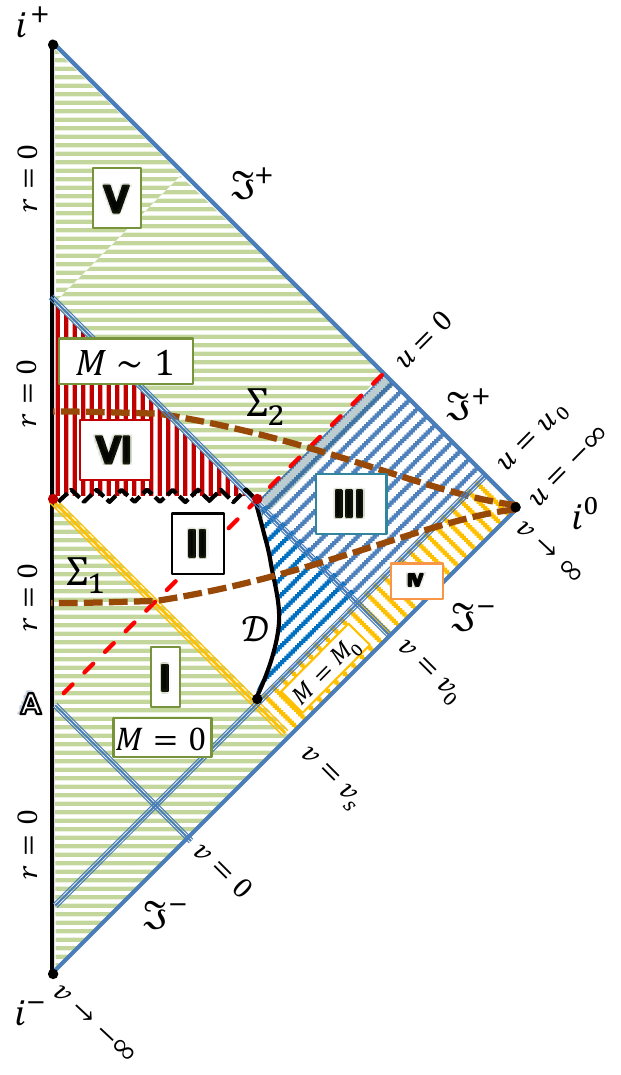}
    \caption{Possible conformal Carter-Penrose diagram associated with the total collapse of a spherically symmetric light shell ($v=v_s$) in a quantum gravity scenario. The spacelike singularity is replaced by a regular (yet super-curved spacetime) region. Regions I, II, III, IV, and V are similar to the corresponding ones in the semiclassical picture (although, {\it stricto-sensu}, there would be no EH). Nevertheless, the collapse would not give rise to a black hole in this scenario since information would leak to the full quantum gravity region VI.  The dashed spacelike hypersurfaces~$\Sigma_1$ and~$\Sigma_2$ represent Cauchy surfaces of this globally hyperbolic spacetime. Information could be preserved along the whole spacetime depending on the equations that evolve the quantum gravity degrees of freedom. Nevertheless, some caveats must be considered, as discussed in Sec.~\ref{sec:paradox}. 
    }
    \label{fig:QG}
\end{figure}

On the other hand, we must acknowledge the possibility that quantum gravity changes the semiclassical picture by resolving the singularity, possibly altering the spacetime of Fig.~\ref{fig:Hiscock1} as, e.g., the one of Fig.~\ref{fig:QG}. By replacing the singularity with a regular, yet super-curved, spacetime region, the quantum-gravity spacetime would gain a full quantum-gravity region, region~VI, but would be deprived of the event horizon (EH).  Still, the quantum-gravity spacetime should preserve the classical and semiclassical (including Hawking radiation) features far from the Planck scale region. Importantly, the spacetime shown in Fig.~\ref{fig:QG} is supposed to be globally hyperbolic in contrast to the one of Figs.~\ref{fig:Hiscock1} and~\ref{fig:Hiscock2} that contain a naked singularity: $\mathbb{S}_2$~\cite{C84}.
 Global hyperbolicity is a much-desired condition for well-posed initial value formulation of physical field equations~\cite{W84}. (Globally-hyperbolic spacetimes $({\cal M}, g_{ab})$ can be foliated by Cauchy surfaces,~$\Sigma_t$, labeled by a real-valued parameter~$t$~\cite{G70}: $({\cal M}, g_{ab})=(\Sigma_{t\in \mathbb{R} }, g_{ab})$.) It is challenging to conceive how information could be conserved in non-globally hyperbolic spacetimes. 

We note, however, that even if quantum gravity turns the spacetime globally hyperbolic as in Fig.~\ref{fig:QG}, it must comply with other nontrivial requirements to avoid loss of information. For instance, assuming that the black hole thermally evaporates until its mass reaches the Planck scale, quantum gravity should allow the storage of an arbitrarily large number of degrees of freedom in a Planck-mass object that would be entangled with the ones of Hawking radiation. Furthermore, the (supposedly) unitary quantum-gravity evolution to the future of the resolved singularity (region~VI; but {\it also} region~V) should lead to a quantum-gravity state ``peaked'' around {\it one} semiclassical geometry (i.e., a classical geometry plus ``small'' fluctuations possibly entangled with matter degrees of freedom). If this assumption---which is far from obvious---turns out to be false, then {\it no} Cauchy surface $\Sigma_2$ to the future of the singularity in a {\it single} semiclassical-geometry realization will contain the full information present in the initial state. In other words, there is no {\it a priori} reason why the full quantum-gravity unitary evolution should preserve the initial-state structure (given at ${\cal J}^-$) of not only being pure but also having the gravity sector ``peaked'' around a (semi)classical geometry---especially when going through a ``deep'' quantum-gravity region such as a (resolved) singularity. 

\begin{figure}[htbp]
    \centering
    \includegraphics[scale=0.80]{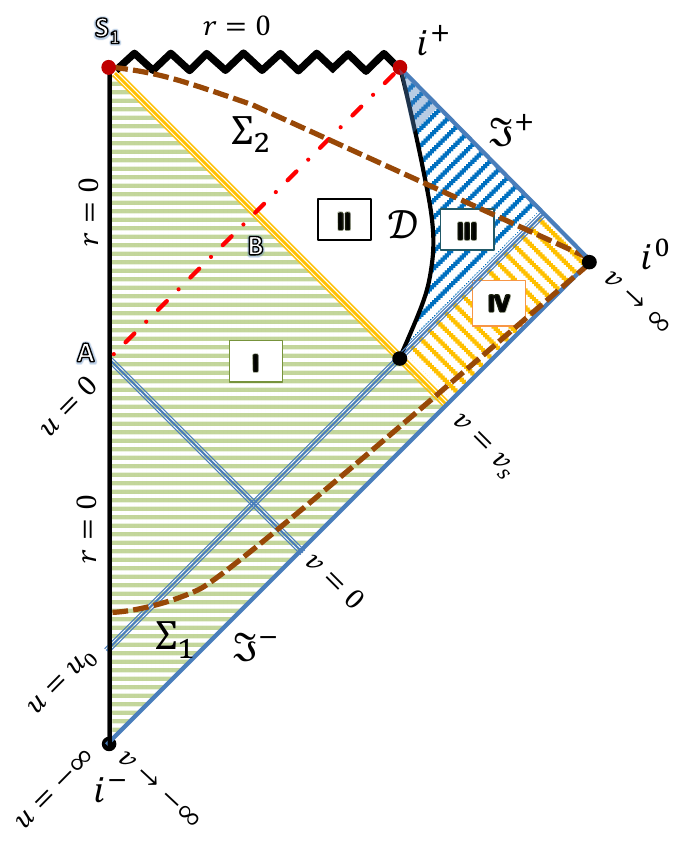}
    \caption{Conformal Carter-Penrose diagram associated with the total collapse of a spherically symmetric light shell ($v=v_s$) with emission of Hawking radiation. This is a globally hyperbolic spacetime but in contrast to the usual semiclassical picture of Figs.~\ref{fig:Hiscock1} and~\ref{fig:Hiscock2}, a black hole remnant is left out at the end.}   
    \label{fig:JC}
\end{figure}

In this paper, we discuss the possibility that the usual semiclassical spacetime of Figs.~\ref{fig:Hiscock1} and~\ref{fig:Hiscock2} is replaced by a globally hyperbolic one  preserving all classical and semiclassical physics far from the Planck region. The simplest possibility would be given by a black hole that evaporates as usual until it reaches the Planck size, after which it would freeze out as shown in Fig.~\ref{fig:JC}. Here, however, we look for a situation where the black hole completely evaporates in addition to being globally hyperbolic (see Figs.~\ref{fig:PLMV1} and~\ref{fig:PLMV2}). To do so, instead of replacing the spacelike singularity of Figs.~\ref{fig:Hiscock1} and~\ref{fig:Hiscock2} with a super-curved spacetime region as was done in Fig.~\ref{fig:QG}, we substitute the naked singularity with a null (thunderbolt) singularity. As a consequence, the corresponding spacetime preserves (i)~the EH null surface, (ii)~all the expected classical and semiclassical features (including Hawking radiation) far from the Planck region, and (iii)~information over the Cauchy surfaces. It must be stressed, however, that because parts of the late-time Cauchy surfaces are inside the EH (see~$\Sigma_2$ in Figs.~\ref{fig:PLMV1} and~\ref{fig:PLMV2}), late-time external observers will only have access to part of the information contained in the asymptotic past (see~$\Sigma_1$ in Figs.~\ref{fig:PLMV1} and~\ref{fig:PLMV2}), vindicating the mixed state of Hawking radiation. This is not any worse than what is found in the more familiar case where a black hole remnant is left at the end, Fig.~\ref{fig:JC}, in which case observers outside the EH can always ``blame'' the black hole for the information they do not have access to. (Nevertheless, in our case, the black hole fully evaporates, as in the usual semiclassical scenario, in the sense that no massive remnant is left over at the end. Note that there are light rays from $\cal{J}^-$ that do not enter the horizon.) Whether Figs.~\ref{fig:PLMV1} and~\ref{fig:PLMV2} of a completely evaporating black hole are correct depends on what quantum gravity will say about the ``last breath'' of a Planck mass black hole.

Instead of considering specific quantum-gravity models (see, e.g., Refs.~\cite{HS93,L93,APR11}), we focus on unveiling general properties of the spacetime given by Figs.~\ref{fig:PLMV1} and~\ref{fig:PLMV2}. (See, e.g., Ref.~\cite{HS10} for other spacetimes investigated irrespective of specific quantum-gravity theories.) We comply with Einstein and semiclassical equations everywhere in the low-curvature regions but not necessarily at the Planck scale, where we only assume the spacetime to possess a Lorentzian metric arbitrarily close to the singularity. This presumes that the forthcoming quantum-gravity theory will provide an effective description of the entire spacetime as a manifold endowed with metric, enabling the discrimination between spacetimes that only differ at the Planck scale.

%

\section{Nonglobally hyperbolic evaporating black hole}
\label{sec:Hiscock}

Let us consider a spherically symmetric thin shell with mass~$M_0$ collapsing at the speed of light~(with light-like coordinate $v=v_s$)~to form a black hole, as shown in Figs.~\ref{fig:Hiscock1} and~\ref{fig:Hiscock2}. We initially consider the usual semiclassical scenario where the black hole completely evaporates with the emission of Hawking radiation, leaving a Minkowski spacetime, region~V. (We shall disregard possible massive particles that enter region~V, making the metric deviate from flatness.) We shall broadly follow Ref.~\cite{Hiscock81} in this section, which will serve to put in context our results of Sec.~\ref{sec:PLMV}.

 The spacetime inside the shell ($v< v_s$), region~I, is flat and the metric can be cast as
\begin{equation}
    ds_{\rm I}^2=-dv^2+2dvdr+r^2d\Omega^2,
\end{equation}
where $r \geq 0$, $d\Omega^2 \equiv d\theta^2 + r^2d\phi^2$, and $v={\rm constant}$ denotes ingoing null geodesics. 

Concerning the metric outside the shell, let us assume that the evaporation of the black hole is negligible for $u< u_0 $, where $u_0={\rm constant}$ represents some outgoing null geodesic. As a consequence, Birkhoff's theorem guarantees that the spacetime in region~IV, defined by $u< u_0 \cap v>v_s$, will be endowed with the usual Schwarzschild vacuum metric:
\begin{equation}
\label{Schwarzschild metric}
    ds_{\rm IV}^2 = -\left(1-\frac{2M_0}{r}  \right)du^2 - 2dudr + r^2d\Omega^2. 
\end{equation}
Then, the evaporation process starts (or becomes important) at $u=u_0$, where the original mass~$M_0$ of the shell begins its gradual conversion into a positive-energy-density outgoing flux of radiation filling region~III (associated with a negative-energy-density ingoing flux in region~II flowing down into the black hole).  Following Hiscock~\cite{Hiscock81}, regions~II and~III are assumed to share a common timelike hypersurface $\mathcal{D}$ (see Figs.~\ref{fig:Hiscock1} and~\ref{fig:Hiscock2}) whose radius {\it monotonically} decreases from some $r=r_0$, at $u=u_0 \cap v=v_s$, to $r=0$, at the naked singularity $\mathbb{S}_2$.

Hence, regions~$\rm{II}$ and~$\rm{III}$ should be well described by the ingoing and outgoing Vaidya metrics
\begin{equation}
\label{ingoing metric}
    ds^2_{\rm II}=-\left(1-\frac{2M_{\rm{II}}(v)}{r}    \right)dv^2+2dvdr + r^2d\Omega^2
\end{equation}
and
\begin{equation}
\label{outgoing metric}
    ds^2_{\rm III}=-\left(1-\frac{2M_{\rm{III}}(u)}{r}    \right)du^2-2dudr + r^2d\Omega^2,
\end{equation}
respectively, where $M_{\rm{II}}(v)$ and $M_{\rm{III}}(u)$ are monotonically decreasing mass functions:
\begin{eqnarray}
    &&  dM_{\rm{II}}(v)/dv<0 \quad (v_s< v <v_0),
    \label{dM/dv}
    \\
    && dM_{\rm{III}}(u)/du<0  \quad (u_0< u <0),
    \label{dM/du}
\end{eqnarray}  
and the coordinate~$r$ must have the same value at every point on the common boundary~$\cal{D}$, as defined in regions~II and~III, to ensure continuity of the 2-sphere areas. The dependence of the mass functions was chosen to reflect that Hawking radiation is dominated by the emission of massless particles, rendering $M_{\rm{III}}$ constant along the $u={\rm constant}$ null rays. An analogous reasoning for the negative-energy-density ingoing flux in region~II drives $M_{\rm{II}}$ to be constant along the $v={\rm constant}$ null rays. Indeed, black holes are not expected to emit non-relativistic particles (say, neutrinos with energies of the order of $1~{\rm eV}$) before their Schwarzschild radius gets as minuscule as $10^{-8}~{\rm m}$. Note, then, that 
\begin{equation}
\label{boundary conditions}
    M_{\rm{II}}(v_s) = M_{\rm{III}}(u_0)=M_0,
    \quad
    M_{\rm{II}}(v_0)= M_{\rm{III}}(0)=0. 
\end{equation}

Now, in order to ensure the correct matching of the Vaidya metrics~(\ref{ingoing metric}) and~(\ref{outgoing metric}) along the boundary of regions~II and~III, we shall demand the first junction condition~\cite{P04}
\begin{equation}
\label{first junction condition}
    h_{ab}^{\rm{II}}  = h_{ab}^{\rm{III}},
\end{equation}
where $h_{ab}$ is the induced metric on the tridimensional hypersurface $\mathcal{D}$. (Because $\mathcal{D}$ is the region where positive- and negative- energy ``particles'' going to ${\cal J}^+$ and entering the black hole are assumed to pop up, respectively, there is no reason to expect the transition between the regions to be smooth on the stress-energy tensor and, thus, to impose the second junction condition $K_{ab}^{\rm{II}}  = K_{ab}^{\rm{III}}$ for the extrinsic curvature $K_{ab}$ on $\mathcal{D}$.) Let $\mathcal{D}$ be covered with coordinates $\xi^i = (\lambda,\theta,\phi)$, where $\partial_\lambda$ is a future-pointing timelike vector. Thus, we can map the $\xi^i$ coordinates to the ones 
defined in regions $\rm{II}$ and $\rm{III}$ on $\mathcal{D}$ as
\begin{equation}
(\lambda, \theta,\phi)  
\stackrel{\mathcal{D}}{\mapsto}
(v (\lambda), r(\lambda), \theta,\phi)
\end{equation}
and
\begin{equation}
(\lambda, \theta,\phi)  
\stackrel{\mathcal{D}}{\mapsto}
(u (\lambda), r(\lambda), \theta,\phi),
\end{equation}
respectively. Hence, the induced metrics are
\begin{equation}
\label{induced metric on II}
    ds^2_{\rm{II}} \stackrel{\mathcal{D}}{=} -\left[\left(1-\frac{2M_{\rm{II}}(v)}{r}\right)\dot{v}^2 -2\dot{v}\dot{r} \right]d\lambda^2 + r^2d\Omega^2
\end{equation}
and
\begin{equation}
\label{induced metric on III}
    ds^2_{\rm{III}} \stackrel{\mathcal{D}}{=} -\left[\left(1-\frac{2M_{\rm{III}}(u)}{r}\right)\dot{u}^2 +2\dot{u}\dot{r} \right]d\lambda^2 + r^2d\Omega^2,
\end{equation}
where a dot represents derivative with respect to $\lambda$. Imposing the first junction condition~(\ref{first junction condition}) gives
\begin{equation}
\label{first junction condition equation}
    \dot{r} \stackrel{\mathcal{D}}{=} \frac{1}{2(\dot{v}+\dot{u})}\left[  \left(1-\frac{2M_{\rm{II}}(v)}{r}\right)\dot{v}^2 - \left(1-\frac{2M_{\rm{III}}(u)}{r}\right)\dot{u}^2   \right].
\end{equation}
Using in the above equation the fact that $\dot{r}<0$ [see discussion below Eq.~(\ref{Schwarzschild metric})] with $\dot{v}>0$ and $\dot{u}>0$ (see Fig.~\ref{fig:Hiscock1}), we get
\begin{equation}
    \label{inequality FJC}
    \left(\frac{1- {2M_{\rm{III}}(u)}/{r}}{1- {2M_{\rm{II}}(v)}/{r}}\right)^{1/2} 
    > \frac{dv}{du}
    >0.
\end{equation}
The $\mathcal{D}$ surface will be timelike if $ds^2<0$ for $\theta,\phi = \rm{constant}$. Imposing this in Eq.~(\ref{induced metric on III}), we obtain
\begin{equation}
   \left(1-\frac{2M_{\rm{III}}(u)}{r}\right)\dot{u}>-2\dot{r}>0,
\end{equation}
because $\dot{r}<0$. Recalling that $\dot{u}>0$, the surface $\mathcal{D}$ will be timelike provided that
\begin{equation}
\label{greater than the AH for III}
    r > 2M_{\rm{III}}(u)
\end{equation}
along $\mathcal{D}$. Applying condition~(\ref{greater than the AH for III}) in Eq.~(\ref{inequality FJC}), we get along $\mathcal{D}$
\begin{equation}
\label{greater than the AH for II}
    r>2M_{\rm{II}}(v).
\end{equation}

Equation~(\ref{greater than the AH for II}) turns out particularly convenient when comparing the relative positions of~$\mathcal{D}$ and the apparent horizon (AH). The latter is given by the geometric location where outgoing null geodesics have vanishing expansion $\theta$. In the coordinates being used, this translates into the geometric location where outgoing null geodesics stem with $dr/dv=0$. Let us make the ansatz (to be confirmed just ahead) that the AH is within region $\rm{II}$. In this case, we consider the equation for outgoing null geodesics ($\theta,\phi={\rm constant}$) given by metric~(\ref{ingoing metric}):
\begin{equation}
\label{null geodesic equation for II}
    \frac{dr}{dv} = \frac{1}{2}\left(1-\frac{2M_{\rm{II}}(v)}{r}   \right).
\end{equation}
By imposing above $dr/dv=0$, we obtain the equation for the AH: 
\begin{equation}
\label{Apparent Horizon equation for region II}
    r(v) \stackrel{AH}{=} 2M_{\rm{II}}(v), \quad v_s < v < v_0.
\end{equation}
By comparing Eqs.~(\ref{greater than the AH for II}) and~(\ref{Apparent Horizon equation for region II}), we see that 
the AH is inside the boundary $\mathcal{D}$ and, thus, dwells in region~$\rm{II}$, confirming the ansatz (see Fig.~\ref{fig:Hiscock2}). Then, the AH pops up at the light shell with $r=2M_0$ [see Eq.~(\ref{boundary conditions})] and diminishes monotonically as ruled by Eq.~(\ref{dM/dv}), eventually vanishing at~$\mathbb{S}_2$. 

We shall also note that the AH 3-surface is timelike. This follows from noticing that the orthogonal vector to the AH hypersurface is spacelike:
\begin{equation}
g^{\alpha\beta}\phi_{,\alpha}\phi_{,\beta}= -4\frac{dM_{\rm{II}}(v)}{dv} > 0,
\end{equation}
where $\phi(x^\mu) \equiv r-2M_{\rm{II}}(v)=0$ defines the~AH. 

We also note that the AH must stay external to the EH since it is everywhere timelike and ends at~$\mathbb{S}_2$. (If it crossed the EH, the AH would not be able to reach $\mathbb{S}_2$ being timelike.) This is in line with the fact that the null energy condition $R_{ab}k^ak^b \geq 0$, for any null vector $k^a$, is not satisfied in region II. For instance, choosing the null vector 
\begin{equation}
    k^\alpha=\left(1, \frac{1}{2}\left(1-\frac{2M_{\rm II}(v)}{r} \right) ,0,0 \right),
\end{equation}
we obtain from Eq.~(\ref{dM/dv}) that
\begin{equation}
    R_{\alpha \beta} k^{\alpha} k^{\beta} = \frac{2}{r^2}\frac{dM_{II}(v)}{dv} < 0.
\end{equation}

Although the precise form of $M_{\rm{II}}(v)$ and $M_{\rm{III}}(u)$ depends on the evaporation details (that we do not address here), we know that the late stage of the evaporation process is dominated by Hawking radiation represented by the translucent strips in Figs.~\ref{fig:Hiscock1} and~\ref{fig:Hiscock2}. Hence, we can give an explicit form for $M_{\rm{III}}(u)$ in this region based on Hawking's prediction. To do so, consider the Hawking temperature as observed by asymptotic static observers for a black hole of mass $M$~\cite{H75}:
\begin{equation}
    \label{Hawking temperature}
    T_H = 1/(8\pi M),
\end{equation}
where it is required that the black hole mass be much larger than the Planck mass: $M \gg 1$. Then, the decrease of~$M$ for a chargeless black hole can be estimated using the Stefan-Boltzmann formula:
\begin{equation}
\label{Stefan-Boltzmann formula}
    {dM}/{dt} = -\sigma AT_H^4,
\end{equation}
where $\sigma\equiv \pi^2 / 60$ is the Stefan-Boltzmann constant (in natural units), $A=16\pi M^2$ is the area of the EH, and $t$ is the proper time of asymptotic static observers. By sticking with the assumption that the radiation reaching the asymptotic future is purely null outgoing [i.e., $M = M_{\rm III} (u)$], we use Eqs.~(\ref{Hawking temperature}) and~(\ref{Stefan-Boltzmann formula}) to write
\begin{equation}
\label{Hawkin prediction}
     \frac{dM_{\rm{III}}(u)}{du} = - \frac{M_{\rm{III}}(u)^{-2}}{15,360}.
\end{equation}
This can be integrated, leading to
\begin{equation}
\label{Outgoing mass}
    M_{\rm{III}}(u) = \left(\frac{-u}{5120}\right)^{1/3},
\end{equation}
where we note that the end of the evaporation here is at $u=0$. We emphasize that Eq.~(\ref{Outgoing mass}) is only a good approximation in the domain of region~III where the optical geometric approximation used to derive Hawking's effect is valid.  

\section{Globally hyperbolic fully evaporating black hole }
\label{sec:PLMV}

\begin{figure}[th]
       \centering
       \includegraphics[scale=0.75]{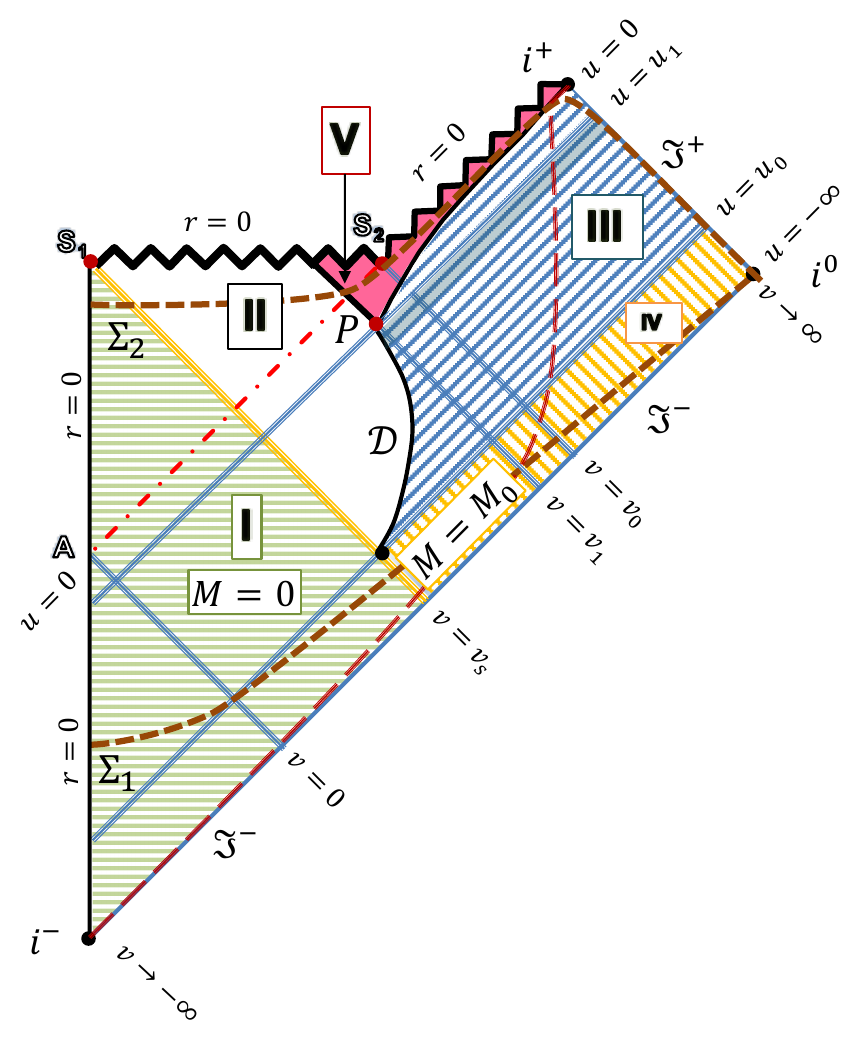}
       \caption{Conformal Carter-Penrose diagram associated with the total collapse of a spherically symmetric light shell into a black hole with mass~$M_0$. This spacetime is similar to the usual semiclassical one except for the quantum-gravity naked singularity that is replaced by a null singularity. The timelike curve~${\cal D}$ separates regions~II and~III ruled by metrics~(\ref{ingoing metric}) and~(\ref{outgoing metric}) of ingoing negative-energy and outgoing positive-energy radiation, respectively. The mass of the black hole is of the order of the Planck mass, $M \sim 1$, at the event $P$. The Planck-scale region near the null singularity is represented by region V that is delimited by the null line $v=v_1$ joining $P$ to the spacelike singularity at $r=0$, and the timelike curve with Planck radius $r = 1$  joining $P$ to $i^+$.  The spacetime is globally hyperbolic, and the dashed spacelike hypersurfaces~$\Sigma_1$ and~$\Sigma_2$ are Cauchy surfaces. Information is preserved on each Cauchy hypersurface, although observers will not have access to the information that enters the black hole. It is worth noting that the black hole will still be held responsible for any information that observers outside the event horizon cannot access.}
       \label{fig:PLMV1}
\end{figure}

\begin{figure}[th]
       \centering
       \includegraphics[width=110mm]{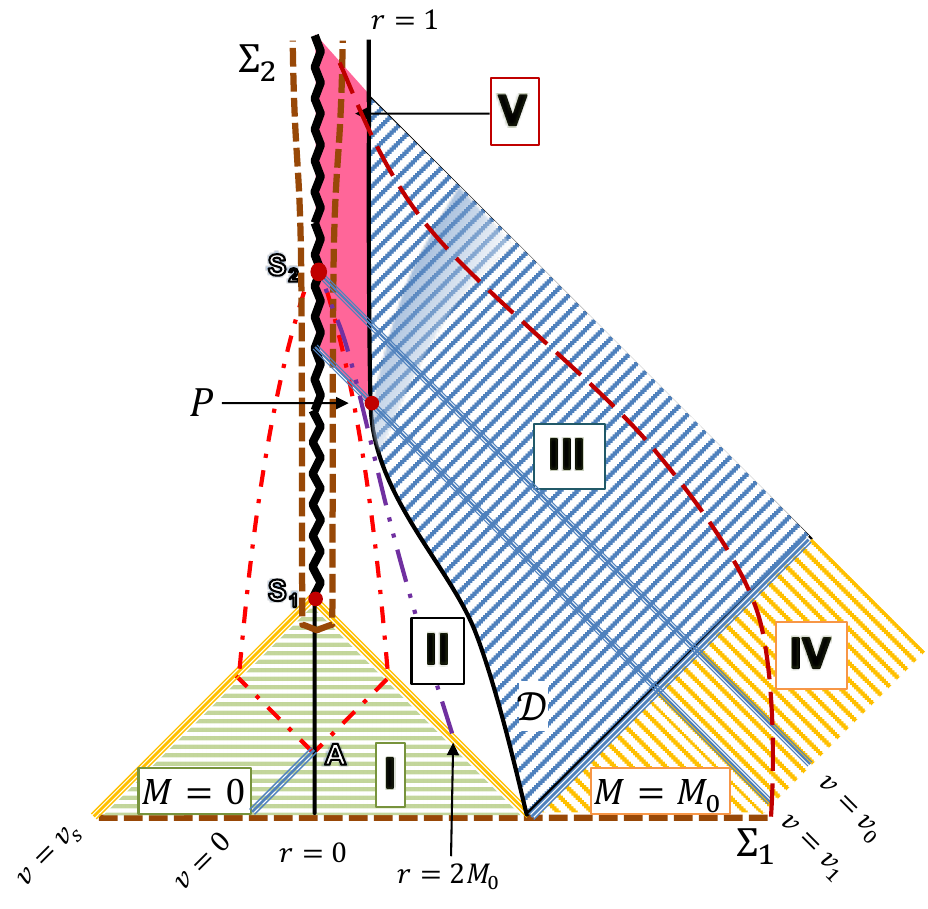}
        \caption{Sketch of the Finkelstein diagram corresponding to the Carter-Penrose one shown in Fig.~\ref{fig:PLMV1}. The diagram is axially symmetric around the $r=0$ axis, but we have avoided repeating information to not overload it. An observer extending from $i^-$ to $i^+$ (as depicted here and in Fig.~\ref{fig:PLMV1} by long-dashed lines) would perceive no particles in regions~I and~IV but this would continuously change in region~III building up into Hawking radiation at $u\lesssim u_1$, eventually. After that, the observer would accuse the presence of a black hole with Planck mass (decreasing as ruled by quantum gravity).  It is particularly illuminating to see how the Cauchy surface~$\Sigma_2$ curves itself, collecting information from past Cauchy surfaces. 
        } 
       \label{fig:PLMV2}
\end{figure}

While the semiclassical theory provides a reliable description of the evaporation process for macroscopic black holes, its predictions are expected to break down when the black hole reaches a mass of the order of the Planck mass. After that, the black hole evaporation is governed by quantum gravity, which would supposedly describe the process in the so-called Planck scale. We shall look for a modification of the spacetime in this regime to make it globally hyperbolic (taking care of not jeopardizing any classical or semiclassical results). The simplest example would be a black hole that evaporates via Hawking radiation until it reaches the Planck size, after which it would freeze out (see Fig. \ref{fig:JC}). Here, however, we look for an instance where the black hole fully evaporates (in the sense that no massive remnant is left over at the end) in addition to being globally hyperbolic, as described by Figs. \ref{fig:PLMV1} and \ref{fig:PLMV2}. (By construction any spacetimes that maintain unaltered classical and semiclassical physics can only be distinguished by their Planck scale different properties.)  

In this token, we shall choose the metric in region~V of Figs.~\ref{fig:PLMV1} and~\ref{fig:PLMV2} ($v> v_1 \cap r\lesssim 1$) to make the spacetime globally hyperbolic as depicted in the diagrams, while the evaporation model discussed in Sec.~\ref{sec:Hiscock} will be valid everywhere outside region $\rm{V}$. Note that~$P$ at $(u, v) = (u_1,v_1)$, representing a 2-sphere with an area of the Planck order, is the meeting point of regions~II, III and~V. It is interesting to note that observers in the far future visiting the asymptotically flat region $ 0 < u < u_1 $ will be influenced (but not governed) by quantum gravity. Hence, we assume that we can also describe this region by metric~(\ref{outgoing metric}) with a general mass function $M_{\rm{III}}(u)$ that should approach the Planck mass,
\begin{equation}
    M_{\rm{III}}(u_1) 
    \sim 1,
\end{equation}
as $u\to u_1$.

Now, let us investigate the conditions that the metric in region~V should satisfy to replace the naked singularity~$\mathbb{S}_2$ in Figs.~\ref{fig:Hiscock1} and~\ref{fig:Hiscock2} by a null (thunderbolt) singularity as in Figs.~\ref{fig:PLMV1} and~\ref{fig:PLMV2}. Previous investigations, e.g. Ref.~\cite{HS93}, have examined the Planck-scale region at the end of black hole evaporation and its connection to thunderbolt singularities based on specific quantum-gravity models. In contrast, we seek to uncover general conditions for the occurrence of a thunderbolt singularity and its prevailing features in a model-independent manner. In this way, let us begin considering a general ingoing Vaidya metric in region~V but, now, with the mass function $M=M_{\rm{V}}(v,r)$ depending on $v$ and $r$:
\begin{equation}
\label{region V metric}
    ds^2_{V} = -\left(1-\frac{2M_{\rm{V}}(v,r)}{r}  \right)dv^2+2dvdr+r^2d\Omega^2,
\end{equation}
where we use the same radial coordinate $r$ to ensure the continuity of the 2-sphere areas. Let us cast $M_{\rm{V}}(v,r)$ as
\begin{equation}
\label{mass function}
M_{\rm{V}}(v,r)=
    \left\{
    \begin{array}{cc}
        \mu_1(v,r), \quad & v \geq v_0,       \\
        \mu_2(v,r), \quad & v_0 \geq v > v_1, \\
    \end{array}
    \right.
\end{equation}
with $\mu_i(v,r)$, $i={1,2}$, being  monotonic positive functions: 
\begin{equation}
\label{monotonic conditions}
    \partial_v \mu_i(v,r)<0,
    \quad
    \partial_r \mu_i(v,r)>0.
\end{equation}
Also, continuity of $M_{\rm{V}}(v,r)$ at $v=v_0$ is ensured by imposing
\begin{equation}
\label{continuity condition}
    \mu_1(v_0,r) = \mu_2(v_0,r).
\end{equation}

The rise of the null singularity depends on choosing Eq.~(\ref{mass function}) such that $r=0$ is identified with an outgoing singular null ray, $u=0$, for $v\geq v_0$. To achieve this, let us consider a one-parameter family of spherically symmetric timelike hypersurfaces~$\mathsf{S}_i$, defined by  $\phi_i(x^\mu) = {\rm constant}$, namely, $r= r_i = {\rm constant}$ for different values of $r_i$, and let us assume that 
$\nabla_\alpha \phi_i \nabla^\alpha \phi_i \stackrel{ r_i \to 0}{\longrightarrow} 0$. Then, in the $r_i\to 0$ limit, $\mathsf{S}_i$ degenerates into a null surface spanned by the null dual vector $\nabla_\alpha r$. 
%
Since $\nabla_\alpha r = (0,1,0,0)$ and 
\begin{equation}
g^{\mu\nu}=
    \left[
    \begin{array}{cccc}
        0 &        1             &     0  &              0                  \\
        1 & 1-2M_{\rm{V}}(v,r)/r &     0  &              0                  \\
        0 &        0             & r^{-2} &              0                  \\
        0 &        0             &     0  & \left(r\sin{\theta}\right)^{-2} \\
    \end{array}
    \right]
\end{equation}
for metric~(\ref{region V metric}), we obtain
\begin{equation}
  \nabla_\alpha r \nabla^\alpha r = 1-{2M_{\rm{V}}(v,r)}/{r}.
\end{equation}
Therefore, a null outgoing ray (labeled by $u=0$) stems in region~V from the evaporation endpoint $\mathbb{S}_2$ at $r=0$ provided  
\begin{equation}
\label{null central curve condition}
    \lim_{r\to 0}~\left[1- {2M_{\rm{V}}(v,r)}/{r}\right] =0. 
\end{equation}
Combining Eqs.~(\ref{null central curve condition}) and~(\ref{mass function}) for $v\geq v_0$ leads us to the conclusion that the existence of a zero-mass null singularity,  $M_{\rm{V}}(v,r) \stackrel{r\to 0}{\longrightarrow} 0$  (for $v\geq v_0$), requires
\begin{equation}
\label{condition on f1}
    \lim_{r\to 0}~{\mu_1(v,r)}/{r} = {1}/{2}.
\end{equation}
(Note that the above requirement is not satisfied, e.g., by linear mass models for evaporating black holes \cite{Hiscock81,Hiscock81a,R06}.)

The singular nature of the outgoing null ray departing from~$\mathbb{S}_2$ at $u=0$ ($r=0$) in region~V ($v \geq v_0$) can be investigated using the Kretschmann scalar $K=R_{abcd}R^{abcd}$:
\begin{equation}
\label{K scalar}
        K =  4r^{-6} 
        \left[
        12 \mu_1^2 - 16 \,r \,\mu_1 \partial_r\mu_1 + 4 r^2 \mu_1 \partial^2_r\mu_1
        + 8r^2 (\partial_r\mu_1)^2 + r^4 (\partial^2_r\mu_1)^2 - 4 r^3 (\partial_r\mu_1) (\partial^2_r\mu_1)
        \right].
\end{equation}
Equation~(\ref{condition on f1}) ensures the divergence of the term $\mu_1(v,r)^2/r^6$ for $r\to 0$ and, consequently, of~$K$ (provided the other terms are not fine-tuned to cancel out this divergence).

Finally, we must impose the mandatory continuity of the induced metrics, $h_{ab}$, on the boundaries of region~$\rm{V}$. (We do not impose the matching of the extrinsic curvatures, as there is no reason to expect the transition of the stress-energy tensor from semiclassical to quantum-gravity regions to be smooth.)
The condition $h_{ab}^{\rm{II}}  = h_{ab}^{\rm{V}}$ is automatically satisfied, since the corresponding induced metrics at $v=v_1$ are
\begin{eqnarray}
    ds_{\rm{II}}^2 && \stackrel{v=v_1}{=}r^2d\Omega^2,
    \\
    ds_{\rm{V}}^2 && \stackrel{v=v_1}{=}r^2d\Omega^2.
\end{eqnarray} 
Concerning the matching between regions III and V, we cover the corresponding boundary at $r=1$ (in natural units)  with coordinates $\xi^i = (\tau,\theta,\phi)$, where $\partial_\tau$ is a future-pointing timelike vector. Thus, we can map accordingly the $\xi^i$ coordinates on the surface to the ones defined in regions III and V as
\begin{eqnarray}
(\tau, \theta,\phi)   
& \stackrel{r=1}{\mapsto}
\ (v (\tau), 1, \theta,\phi),
\\
(\tau, \theta,\phi)  
& \stackrel{r=1}{\mapsto}
 \ (u (\tau), 1, \theta,\phi).
\end{eqnarray}
In this way, the induced metrics are
\begin{eqnarray}
    ds_{\rm{III}}^2 \stackrel{r=1}{=}-\left[1- 2 M_{\rm{III}}(u)\right]\left(\frac{du}{d\tau}\right)^2d\tau^2+ d\Omega^2
\end{eqnarray}
and
\begin{eqnarray}
    ds_{\rm{V}}^2 \stackrel{r=1}{=}-\left[1- 2 M_{\rm{V}}(v,1)\right]\left(\frac{dv}{d\tau}\right)^2d\tau^2+ d\Omega^2.
\end{eqnarray}
Then, the junction condition, $h_{ab}^{\rm{III}}  = h_{ab}^{\rm{V}}$, reads
\begin{equation}
    \frac{du}{dv}\stackrel{r=1}{=}\left(\frac{1-2M_{\rm{V}}(v,1)}{1-2M_{\rm{III}}(u)}\right)^{1/2},
    \label{MIII_MV}
\end{equation}
where we impose $M_{\rm III} , M_{\rm V} \lesssim 1/2$ at the event $P$, $(u,v)=(u_1,v_1)$. This, combined with~$\partial_v M_{\rm V} (v,1)<0$, $\partial_u M_{\rm III} (u) < 0$, is enough to guarantee that Eq.~(\ref{MIII_MV}) is well behaved on the whole boundary.

Let us now investigate the behavior of the AH in region $\rm{V}$. The equation for outgoing radial null geodesics assuming the line element~(\ref{region V metric}) is
\begin{equation}
\label{null geodesic equation for V}
    \frac{dr}{dv} = \frac{1}{2}\left(1-\frac{2M_{\rm{V}}(v,r)}{r}   \right).
\end{equation}
The implicit equation for the AH is obtained by imposing $dr/dv=0$ in Eq.~(\ref{null geodesic equation for V}): 
\begin{equation}
    \label{Apparent Horizon equation}
    r(v)\stackrel{AH}{=} 2M_{\rm{V}}[v,r(v)].
\end{equation}
We note that Eq.~(\ref{Apparent Horizon equation}) depends on the $M_{\rm V}(v,r)$ function that is presumably fixed by quantum gravity. For the sector $v_1< v\leq v_0$ in region V, Eq.~(\ref{Apparent Horizon equation}) reads
\begin{equation}
\label{AH mu2}
    r(v) \stackrel{AH}{=}2\mu_2[v,r(v)] \quad (v_1 < v \leq v_0).
\end{equation}
We shall choose the monotonically decreasing function $\mu_2(v,r)$, see Eq.~(\ref{monotonic conditions}), to vanish at the evaporation endpoint $\mathbb{S}_2$. It follows from Eq.~(\ref{AH mu2}), then, that the AH has endpoint at $\mathbb{S}_2$, as in the usual semiclassical picture. (Compare Figs.~\ref{fig:Hiscock2} and~\ref{fig:PLMV2}.)

We shall also note that the AH will be timelike (as in the usual semiclassical case), provided
\begin{equation}
\label{timelike Ah character}
    \partial_rM_{\rm{V}}(v,r)<{1}/{2}.
\end{equation}
This can be seen by calculating the norm of the orthogonal vector to the AH hypersurface 
($\phi^{\rm{V}} (x^\mu) \equiv r-2M_{\rm{V}}(v,r)=0$):
\begin{equation}\label{aux}
g^{\alpha\beta} \phi^{\rm{V}}_{\; ,\;\alpha} \phi^{\rm{V}}_{\; ,\;\beta}= -4\partial_vM_{\rm{V}}(v,r)\left[1-2\partial_rM_{\rm{V}}(v,r)\right],
\end{equation}
and using $\partial_v M_{\rm{V}}(v,r)<0$ [see Eq.~(\ref{monotonic conditions})] to conclude that the left-hand side of Eq.~(\ref{aux}) is positive when Eq.~(\ref{timelike Ah character}) is respected.
\begin{figure}[t]
       \centering
       \includegraphics[width=120mm]{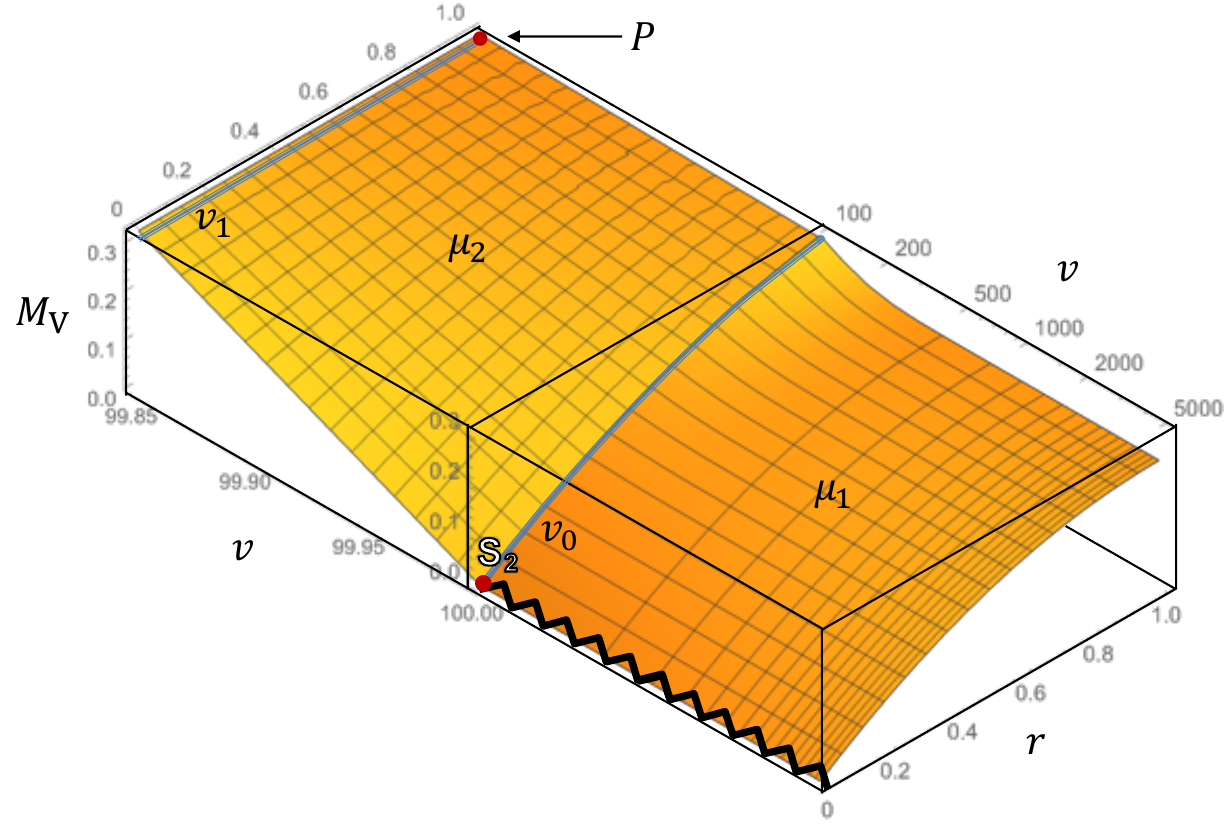}
        \caption{
        Plot of $M_V(v,r)$ given by Eqs.~(\ref{f1}) and~(\ref{f2}) assuming a light shell with $M_0=100$ (in natural units) falling down with $v=v_s=50$. The singularity~$\mathbb{S}_2$ is at $v=v_0=100$. Here, $v_1=99.85$. Note that the scale used for the $v$ axis in the~$\mu_1$ and~$\mu_2$ sectors are distinct.
        } 
       \label{fig:mu1mu2}
\end{figure}

In order to keep the spacetime as close as possible to the usual semiclassical one, we wish to locate the~AH external to the~EH. A necessary condition for this is that $R_{ab} k^a k^b < 0$ for some null vector $k^a$~\cite{W84}.  The Ricci scalar in region~V for $v_0 \geq  v > v_1$ reads
\begin{eqnarray}
\label{Ricci scalar v<v_0}
    R_{\alpha \beta} 
    &=&
    \frac{\partial^2_r\mu_2(v,r)}{r}\delta_\alpha^v\delta_\beta^r - r\partial_r^2\mu_2(v,r)\delta_\alpha^\theta\delta_\beta^\theta
    \nonumber \\
    &+&\frac{2 \partial_v\mu_2(v,r)-\left(r-2 \mu_2(v,r)\right) \partial^2_r\mu_2(v,r)}{r^2}\delta_\alpha^v\delta_\beta^v
    \nonumber \\
    &-&r\partial_r^2\mu_2(v,r)(\sin{\theta})^2\delta_\alpha^\phi\delta_\beta^\phi.
\end{eqnarray} 
The contraction of $R_{\alpha \beta}$ with the null vector
\begin{equation}
\label{null vector example}
    k^\alpha = \left(1,\frac{1}{2}\left(1-\frac{2\mu_2(v,r)}{r} \right),0,0    \right)
\end{equation}
gives $R_{\alpha\beta} k^\alpha k^\beta < 0$ provided
\begin{equation}
\label{wald preposition with k}
    \frac{\partial_r^2\mu_2(v,r)}{r}\left(\frac{\mu_2(v,r)}{r} -\frac{1}{2}\right) 
    <
    -\frac{2\partial_v\mu_2(v,r)}{r^2}.
\end{equation}

Let us now select a particular mass function $M_{\rm V}(v,r)$ to realize the metric of region~V of our globally hyperbolic spacetime containing the thunderbolt singularity (see Fig.~\ref{fig:mu1mu2}). In natural units (where $r=1$ corresponds to the Planck length), let us choose 
\begin{equation}
\label{f1}
    \mu_1(v,r) = \frac{r}{2}\left[ 1-\frac{r}{2}\tanh\left( \frac{v}{v_0} \right) \right],
\end{equation}
\begin{equation}
\label{f2}
    \mu_2(v,r) 
      = 
      \mu_1(v,r) \left(1-\frac{\tanh{(1-v/v_0)}}{\tanh{(1-v_1/v_0)}}\right)
      +
      M_0\frac{\tanh{(1-v/v_0)}}{\tanh{(1-v_s/v_0)}},
\end{equation}
where the light shell has $M_0=100$ falling down freely at $v=v_s=50$. The singularity~$\mathbb{S}_2$ is at $v=v_0=100$. Here,  $v_1=99.85$.  

It is already clear from Fig.~\ref{fig:mu1mu2} that  $\partial_r \mu_i (v,r)>0\; (i=1,2)$, satisfying the second condition of Eq.~(\ref{monotonic conditions}). Concerning the first condition, it is also clear from Fig.~\ref{fig:mu1mu2} that $\partial_v \mu_2 (v,r)<0$, while that $\partial_v \mu_1 (v,r)<0$ is less obvious but can be shown true via a direct calculation:
\begin{equation}
\label{partial_v mu1}
        \partial_v\mu_1(v,r) = -\frac{r^2 \left[ {\rm sech} \left(v/v_0\right)\right]^2}{4v_0} <0.
\end{equation}
We also note that the continuity condition~(\ref{continuity condition}) and the zero-mass null singularity constraint~(\ref{condition on f1}) are trivially satisfied by Eqs.~(\ref{f1}) and~(\ref{f2}). Concerning the condition demanded below Eq.~(\ref{MIII_MV}) that $M_{\rm V}(v,r) \lesssim 1/2$ at the event $P$, one can see that this is also respected from Fig.~\ref{fig:mu1mu2}, namely, $M_{\rm V}(v_1,1) = 1/3$.

Next, let us evaluate the Kretschmann scalar in region~V. For the sector $v \geq v_0$, we use $M_{\rm V}(v,r)=\mu_1 (v,r)$ obtaining
\begin{equation}
    K_1 = \frac{2}{r^4}\left(r^2\left(\tanh\left(\frac{v}{v_0}\right)\right)^2 -2r\tanh\left(\frac{v}{v_0}\right)+2   \right),
\end{equation}
while in the sector $v_0 \geq v > v_1$,  we use $M_{\rm V}(v,r)=\mu_2 (v,r)$, getting
    \begin{eqnarray}
        K_2
        & = &
        \frac{2}{r^6} 
        \left[
              24M_0^2 \left(\frac{\tanh (1-v/v_0)}{\tanh (1-v_s/v_0)}\right)^2
              -2 r \frac{\tanh (1-v/v_0)}{\tanh (1-v_1/v_0)}
              \left[
                  4 M_0 \frac{\tanh (1-v/v_0)}{\tanh (1-v_s/v_0 )}          
              \right.            
       \right.          
   \nonumber \\ 
        & + & 
             \left.
        r^3 \left(\tanh \left( \frac{v}{v_0} \right) \right)^2           
           - 2 r^2 \tanh \left( \frac{v}{v_0} \right) 
                          + 2 r 
            \right]
        + 8 M_0 r \frac{\tanh (1-v/v_0)}{\tanh (1-v_s/v_0)}                 
   \nonumber \\ 
       & + & 
       \left.
       r^2 \left( 1+  \left(\frac{\tanh (1-v/v_0)}{\tanh (1-v_s/v_0)}\right)^2 \right)
        \left(
              r^2 \left(\tanh \left( \frac{v}{v_0} \right)  \right)^2   
        -             
        2 r \tanh \left(\frac{v}{v_0}\right)+2 
       \right) 
       \right].
\end{eqnarray}
The divergence of $ K_1 (v,r) $  as $r\to 0$ reflects the singular nature of the outgoing thunderbolt ray, $u=0$ ($v\geq v_0$), while the divergence of $K_2 (v,r)$ as $r\to 0$ stands for the fact that the hypersurface $r={\rm constant}\to 0$ inside the black hole remains singular, as usual.

Finally, we shall discuss the properties of the AH in the region V. Firstly, let us note that 
\begin{eqnarray}
\label{partial r f2}
    \partial_r\mu_2(v,r) &=& \frac{1}{2}
                           \left[
                                1-\frac{\tanh \left(1-v/v_0\right)}{\tanh
                                \left(1-v_1/v_0\right)}
                            \right]
                            \left[
                                1- r \tanh \left(\frac{v}{v_0}\right)
                            \right]
                            \nonumber \\
                        &<& \frac{1}{2} \quad  (v_0> v > v_1),                         
\end{eqnarray}
satisfying condition~(\ref{timelike Ah character}) for the AH to be timelike.
Besides, an analysis of Eq.~(\ref{wald preposition with k}) shows that $R_{ab}k^ak^b<0$ and, thus, the AH may lie outside the EH, as we will see to be the case here. 
\begin{figure}[th]
       \centering
       \includegraphics[width=120mm]{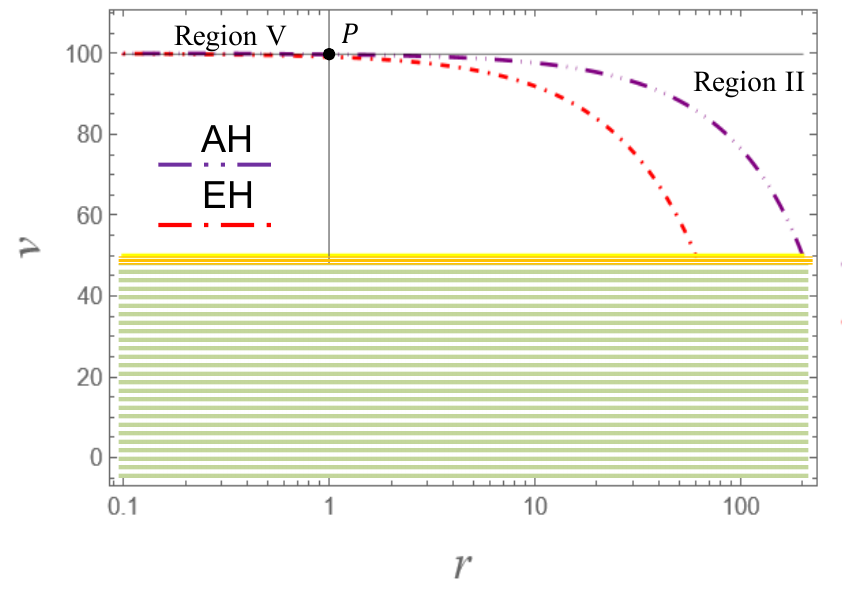}
        \caption{Numerical evolution for the AH and EH assuming a light shell with $M_0=100$ (in natural units). Both the horizons harmoniously approach the singularity at $v=v_0=100$. They are plotted outside the light shell set at $v=v_s=50$ (compare with Figs.~\ref{fig:PLMV1} and~\ref{fig:PLMV2}). The event $P$ is at $v_1 = 99.85$.  
        } 
       \label{fig:EH_AH}
\end{figure}
In order to plot the AH and EH with our choice of $M_V(v,r)$ given in Eq.~(\ref{mass function}) supplemented by Eqs.~(\ref{f1}) and~(\ref{f2}), let us impose the second junction condition between regions~II and~V (i.e., $K_{ab}^{\rm II}=K_{ab}^{\rm V}$ on the corresponding boundary for the extrinsic curvature $K_{ab}$), leading to 
\begin{equation} \label{MII=MV}
    M_{\rm{II}}(v_1) = M_{\rm{V}}(v_1,r).
\end{equation}
Next, let us choose $M_{\rm{II}}(v)$ as
\begin{equation}
\label{mass function in II}
    M_{\rm{II}}(v) =M_0 \frac{\tanh(1-v/v_0)}{\tanh(1-v_s/v_0)},
\end{equation}
which satisfies Eqs.~(\ref{dM/dv}), (\ref{boundary conditions}), and~(\ref{MII=MV}). As a result, the AH in region~II is given by [see Eq.~(\ref{Apparent Horizon equation for region II})]
\begin{equation}
\label{AH II}
    r(v)\stackrel{AH}{=} 2M_0\frac{\tanh(1-v/v_0)}{\tanh(1-v_s/v_0)}.
\end{equation}
Using Eq.~(\ref{AH II}) and Eq.~(\ref{AH mu2}) [with the mass~(\ref{f2})], one can numerically plot the AH in the regions~II and~V, respectively (see Fig.~\ref{fig:EH_AH}). Moreover, the plot of the EH is given by numerically solving the null geodesic equations~(\ref{null geodesic equation for II}) and~(\ref{null geodesic equation for V}) for masses~(\ref{mass function in II}) and~(\ref{f2}) with the boundary condition $r(v_0)=0$, since the EH disappears at the evaporation endpoint~$\mathbb{S}_2$ (see Fig. \ref{fig:EH_AH} and note the mono-log scale used to make explicit the different behavior of the EH and AH in regions $\rm{II}$ and $\rm{V}$). It is clear from the graphic that the AH lies outside the EH and both horizons end at $\mathbb{S}_2$, as physically expected.

   \section{Black holes not as ordinary black boxes}
   \label{Page}

The goal of this section is to clarify the distinction between the two prevailing issues concerning the black hole information loss: 
\begin{itemize}
    \item [A.] the information loss observed at the final stages of the evaporation due to the black hole internal singularity expected to eventually erase information that enters the horizon, and
   \item [B.] the paradox that would arise (much earlier) in the semiclassical regime (after the Page time) if one imposes that black holes must behave as any other ordinary thermodynamical system. 
\end{itemize}

Global hyperbolicity of the spacetime of Fig.~\ref{fig:PLMV1} and the ensuing conservation of information at each Cauchy surface may not be enough to make supporters of the ``central dogma''~\cite{Almheiri21} feel comfortable since they would like all this information to be accessible to external observers. In fact, it has been argued that if information that falls into the black hole does not begin to be accessible to external observers after some given (Page) time, a paradoxical situation would arise in the semiclassical regime when the black hole would still be macroscopic. The argument goes as follows (see, e.g., Ref.~\cite{Almheiri21}).

Let $\Sigma_\zeta$ be a spacelike foliation of region ${\rm I}\cup {\rm II}\cup {\rm III}\cup {\rm IV}$ (of any of Figs.~\ref{fig:Hiscock1}-\ref{fig:PLMV2}) with $\zeta\in \mathbb{R}$ being a labeling increasing to the future. Now, let $\Sigma_\zeta^{in}$ be the intersection of $\Sigma_\zeta$ with the black hole and $\Sigma_\zeta^{out}\equiv \Sigma_\zeta \backslash  \Sigma_\zeta^{in}$. Then, tracing out the degrees of freedom of the quantum fields in the region $\Sigma_\zeta^{in}$ [resp., $\Sigma_\zeta^{out}$] leads to the reduced fields' state $\hat{\rho}_{out}(\zeta )$ [resp., $\hat{\rho}_{in}(\zeta )$], whose corresponding von Neumann entropy we call $S^{vN}_{out}(\zeta)$ [resp., $S^{vN}_{in}(\zeta)$]. It is  well known that if the state of the fields on $\Sigma_\zeta$ is pure (hence, with vanishing von Neumann entropy), then
\begin{eqnarray}
S^{vN}_{out}(\zeta)=S^{vN}_{in}(\zeta),
    \label{eq:SvNinout}
\end{eqnarray} 
after proper regularization
and renormalization.\footnote{For free fields in $d$ spacetime dimensions, the von Neumann entropy for general Hadamard states associated with a (spatial) region $\Sigma$ with boundary $\partial \Sigma$ can usually be cast as $$S_{\Sigma}^{vN}=\sum_{n<d-2}\frac{f_n(\partial \Sigma)}{\epsilon^{d-2-n}} + f(\partial\Sigma) \log \epsilon+S_0\left(\Sigma\right),$$ where $\epsilon$ is a short-distance cutoff, $n=0,1,\cdots $ are positive integers, $f$ and $f_n$ are state-independent constants (although the $f_n$'s depend on the regularization process chosen) expressed as integrals on $\partial \Sigma$, and $S_{0}(\Sigma)$ is a finite state-dependent term. In $d=4$, $f_0(\partial \Sigma)\propto A(\partial \Sigma)$, where $A(\partial \Sigma)$ is the area of $\partial \Sigma$, giving an area law for the leading term of von Neumann entropy divergences. If one is interested in variations of the entropy, one can renormalize $S_{\Sigma}^{vN}$ by subtracting its value with respect to some reference (Hadamard) state. For more details (including interacting theories), see Refs.~\cite{CH22, Solodukhin11, BFLW16, Miqueleto21} and references therein.} 

Now, the assumption that Hawking radiation does {\it not} encode information that falls into the black hole implies that
\begin{eqnarray}
S_{out}^{vN}(\zeta )~\textrm{is a {\it strictly increasing} function of}~\zeta.
    \label{eq:SvNincreasing}
\end{eqnarray}
In particular, if we assume that Hawking radiation is truly a thermal state at Hawking temperature $T_H$, then $S^{vN}_{out}(\zeta)$ should be able to be evaluated from the same (black hole) parameters that determine $T_H$.


On the other hand, {\it if} one {\it associates} the thermodynamical entropy of the black hole, $S^{\rm therm}_{bh}$, as due to $\hat{\rho}_{in} (\zeta)$ at $\Sigma_\zeta^{in}$, we would have 
\begin{equation}
S_{in}^{vN}(\zeta) 
\leq 
\left. S_{in}^{vN}[\tilde{\rho}]\right|_{\rm sup}
\longleftrightarrow
S^{\rm therm}_{bh}(\zeta) 
= A(\zeta)/4,
    \label{eq:SvNBH}
\end{equation}
where the first inequality is the trivial statement that $S_{in}^{vN}$ is limited by the supremum of the von Neumann entropy among all possible states $\tilde{\rho}$ (satisfying the relevant coarse-grained constraints), and `$\longleftrightarrow$' stands for the {\it assumption} that 
$\left. S_{in}^{vN}[\tilde{\rho}]\right|_{\rm sup}$ 
equals the thermodynamical entropy of the black hole given by the Bekenstein-Hawking formula with $A(\zeta)$ being the area of the event horizon on $\Sigma_\zeta$.

It is quite obvious that Eqs.~(\ref{eq:SvNinout}), (\ref{eq:SvNincreasing}), and (\ref{eq:SvNBH}) are in tension: a strictly increasing function $S_{out}^{vN}(\zeta)$ bounded from above by the decreasing function $A(\zeta)/4$. This is the essence of the reasoning
used by the followers of the ``central dogma'' to argue that Eq.~(\ref{eq:SvNincreasing}) should be violated after some (Page) time; i.e., that Hawking radiation should eventually carry away information that fell into the black hole, ``purifying'' $\hat{\rho}_{out}(\zeta)$ for sufficiently large values of $\zeta$. Yet, subtle as it may be, the identification in Eq.~(\ref{eq:SvNBH}) depends on assuming that black holes are like any common (``black box'') thermodynamical system, 
which is a quite nontrivial assumption.


While the walls of a ``black box'' do not have to ``react'' to attempts of external observers to ``hide'' entropy in its interior, the event horizon (being a {\it dynamical} entity) ``denounces'' any such attempt by changing its area. Therefore, while the entropy of a ``black box''  {\it must be} due to whatever it contains (to ensure a meaningful second law of thermodynamics), this is not so for black holes (being consistent with the {\it standard} view that modes falling into the black hole are inaccessible to external observers). According to the standard semiclassical description, the event horizon should be seen as a {\it causal} boundary separating degrees of freedom that (i)~escape to the future null infinity and account for Hawking radiation from those that (ii)~fall into the black hole (being both entangled with each other). In this way, the tension among Eqs.~(\ref{eq:SvNinout}), (\ref{eq:SvNincreasing}), and (\ref{eq:SvNBH}) is avoided, not due to information ``leaking'' from the black hole [i.e., violation of Eq.~(\ref{eq:SvNincreasing})], but because the identification in Eq.~(\ref{eq:SvNBH}) would not be valid.

Indeed, in the semiclassical realm, Eq.~(\ref{eq:SvNincreasing}) is necessary to maintain the generalized second law true. This can be already seen from a simple calculation. Assuming that Hawking radiation as measured by asymptotic Killing observers is thermal with temperature $T_{H}$ as given by Eq.~(\ref{Hawking temperature}), each radiation energy element $dM_{H}$ carries an entropy element $dS_{H}=(4/3) \times dM_{H}/T_H $. This increase of external-to-the-horizon entropy is accompanied by a decrease of the black hole thermodynamical entropy of $dS_{bh}^{\rm therm}=-8\pi M dM$, leading to
\begin{equation}
\left|\frac{dS_{H}}{dS_{bh}^{\rm therm}}\right|=\frac{4}{3}>1.
\end{equation}
A more detailed analysis taking into account the full density matrix  $\hat{\rho}_{out}$ describing Hawking radiation (including grey-body factors) leads to \cite{Zurek82} 
\begin{equation}
\left|\frac{dS_{H}}{dS_{bh}^{\rm therm}}\right|\equiv R \gtrsim 1,
\end{equation}
which ensures the validity of the generalized second law of thermodynamics.

In summary, if one adopts the ``central dogma'' and assumes that black holes must behave as any other ordinary thermal system, Hawking radiation should begin to deviate from thermality still in the semiclassical domain (after some Page time), which is considered paradoxical for being at odds with the semiclassical derivation of the Hawking effect. On the other hand, if one does not adopt the ``central dogma,'' the black-hole degrees of freedom (and thus its Hilbert space), do not need to be associated with the modes that fell into the black hole. In this case, black holes (with their horizons) must be seen as independent dynamical entities with their own Hilbert spaces, existing no contradiction between $A/4$ being the logarithm of the dimension of the black-hole Hilbert space (e.g., representing qubits encoded in each quantum of area) and the much larger Hilbert space associated with the swallowed matter/radiation.

   \section{Discussion and closing remarks }
   \label{Final}

The black hole formation and subsequent evaporation into thermal radiation, as suggested by semiclassical arguments, do not violate any physical principles, even if most information is eventually lost~\cite{UW17}. Nevertheless, quantum gravity may modify the semiclassical picture depending on its impact on the Planck scale physics (Fig.~\ref{fig:QG}). We have discussed here the case where the final naked singularity of the usual semiclassical spacetime, Figs.~\ref{fig:Hiscock1} and~\ref{fig:Hiscock2}, is replaced by a null (thunderbolt) singularity extending to $i^+$, Figs.~\ref{fig:PLMV1} and~\ref{fig:PLMV2}. All classical and semiclassical features are preserved in small curvature regions. In particular, late-time observers will experience Hawking radiation at $u\lesssim u_1$, after which they accuse the presence of a Planck-mass black hole (decreasing as ruled by quantum gravity). A pleasant feature of this spacetime is that it is globally hyperbolic. As a result, information is preserved on each Cauchy hypersurface, where the evolution is assumed to be unitary. Despite this, the fact that late-time Cauchy surfaces ``invade'' the EH prevents late-time observers from accessing complete early-time information.  Thus, even after late-time observers (those ending at $i^+$) do not see Hawking radiation anymore (see subset $u_1\lesssim u \lesssim 0$ in region~III of Fig.~\ref{fig:PLMV1}), they can still ``blame'' the hole for the thermal state of Hawking radiation since its degrees of freedom will be still entangled with the ones inside the hole. It is worth emphasizing that some of the seemingly \textit{ad hoc} choices we have made here (e.g., specific mass functions) were merely intended to provide a concrete realization of the globally hyperbolic evaporation scenario. Without a full quantum gravity theory, nothing can be said about the robustness of the conditions that lead to this scenario, neither in favor nor against it. Yet, we believe that the existence of a fully evaporating black hole scenario where information and unitarity can be preserved (with ``minimal'' deviations from the semiclassical picture) is worthy of attention.

\ack{
   One of us (GM) would like to acknowledge various conversations with Bill Unruh and Bob Wald on the information loss in black holes over the years. The authors also thank Arderucio-Costa for profitable discussions regarding the Page time and for calling our attention to Ref.~\cite{Arderucio24}. J.~P.\ and D.~V.\ were fully and partially supported by Sao Paulo Research Foundation (FAPESP) under grants 2022/14028-3 and 2023/04827-9, respectively. G.~M.  acknowledges partial support from FAPESP and Conselho Nacional de Desenvolvimento Cientifico e Tecnologico under grants 2022/10561-9 and 301508/2022-4, respectively. D.~V. would like to thank also the Institute for Quantum Optics and Quantum Information of the Austrian Academy of Science for hosting him for the sabbatical year.}


\section*{References}
\begin{thebibliography}{99}

\bibitem{M09}
S. D. Mathur,
{ \it The information paradox: A pedagogical introduction,}
{\it Class. Quant. Grav.} {\bf 26} (2009) 224001.

\bibitem{Parikh00}
M. K. Parikh and F. Wilczek
{\it Hawking radiation as tunneling,}
{\it Phys. Rev. Lett.} {\bf 85} (2000) 5042.


\bibitem{Zhang09}
B. Zhang, Q.-Y. Cai, L. You, and M.-S. Zhan,
{\it Hidden messenger revealed in Hawking radiation: A resolution to the paradox of black hole information loss,}
{\it Phys. Lett. B} {\bf 675} (2009) 98.


\bibitem{Almheiri21}
A. Almheiri, T. Hartman, J, Maldacena, E. Shaghoulian, and A. Tajdini,
{\it The entropy of Hawking radiation,}
{\it Rev. Mod. Phys.} {\bf 93} (2021) 035002.

\bibitem{H76}
S. W. Hawking, 
{\it Breakdown of predictability in gravitational collapse,}
{\it Phys. Rev. D} {\bf 14} (1976) 2460.

\bibitem{W94book}
R. M. Wald,
{\it Quantum Field Theory in Curved Spacetime and Black Hole Thermodynamics,}
Chicago: University of Chicago Press (1994).

\bibitem{UW95}
W. G. Unruh and R. M. Wald, 
{\it On evolution laws taking pure states to mixed states in quantum field theory.}
{\it Phys. Rev. D} {\bf 52} (1995) 2176.

\bibitem{UW17}
W. G. Unruh and R. M. Wald, 
{\it Information loss,}
{\it Rep. Prog. Phys.} {\bf 80} (2017) 092002.

\bibitem{Gan20}
W.-C. Gan and F.-W. Shu,
{\it Information loss paradox revisited: Farewell firewall?}
{\it Int. J. Mod. Phys. D} {\bf 29} (2020) 2043019.

\bibitem{P04book}
R. Penrose,
{\it The Road to Reality: a Complete Guide to the Laws of
the Universe,}
London: Jonathon Cape (2004).

\bibitem{W18}
D. Wallace,
{\it Why Black Hole Information Loss is Paradoxical,}
arXiv:1710.03783.

\bibitem{P93}
D. N.  Page, {\it Information in black hole radiation}, 
Phys. Rev. Lett.\ {\bf 71} (1993) 3743.

\bibitem{P13}
D. N.  Page, {\it Time dependence of Hawking radiation entropy}, 
J. Cosmol. Astropart. Phys {\bf 09} (2013) 028.

\bibitem{C84}
R. M. Wald, 
{\it Black holes, singularities, and predictability,}
in {\it ``Quantum Theory of Gravity -- Essays in honor of the $60^{\rm th}$ birthday of B. S. DeWitt,}
Bristol: Adam Hilger (1984).

\bibitem{G70}
R. P. Geroch, 
{\it Domain of dependence,}
{\it J. Math. Phys.} {\bf 11} (1970) 437.

\bibitem{W84}
R. M. Wald, 
{\it General Relativity,}
Chicago: University of Chicago (1984).

\bibitem{HS93}
S. W. Hawking and J. M. Stewart,
{\it Naked and thunderbolt singularities in black hole evaporation,}
{\it Nuc. Phys. B} {\bf 400} (1993) 393.

\bibitem{L93}
D. A. Lowe, 
{\it Semiclassical approach to black hole evaporation,}
{\it Phys. Rev. D} {\bf 47} (1993) 2446.

\bibitem{APR11}
A. Ashtekar, F. Pretorius, and F. M. Ramazano\ifmmode \breve{g}\else \u{g}\fi{}lu, 
{\it Evaporation of two-dimensional black holes,} 
{\it Phys. Rev. D} {\bf 83} (2011) 044040.

\bibitem{HS10}
S. Hossenfelder and L. Smolin,
{\it Conservative solutions to the black hole information problem,}
{\it Phys. Rev. D} {\bf 81} (2010) 064009.

\bibitem{Hiscock81}
W. A. Hiscock,
{\it Models of evaporating black holes. II. Effects of the outgoing created radiation,}
{\it Phys. Rev. D} {\bf 23} (1981) 2823.

\bibitem{P04}
E. Poisson, 
{\it A Relativist's Toolkit: The Mathematics of Black-Hole Mechanics,}
Cambridge: University of Cambridge (2004).

\bibitem{H75}
S. W. Hawking, 
{\it Particle creation by black hole,}
{\it Commun. Math. Phys.} {\bf 43} (1975) 199.

\bibitem{R06}
D. W. Ring,
{\it A linear approximation to black hole evaporation,}
{\it Classical and Quantum Gravity} {\bf 23} (2006) 5027.

\bibitem{Hiscock81a}
W. A. Hiscock,
{\it Models of evaporating black holes. I,}
{\it Phys. Rev. D} {\bf 23} (1981) 2813.

\bibitem{CH22} 
H. Casini and M. Huerta, 
{\it Lectures on entanglement in quantum field theory,}
{\it PoS} {\bf TASI2021} (2022) 2
(arXiv:2201.13310).

\bibitem{Solodukhin11}
S. N. Solodukhin, 
{\it Entanglement entropy of black holes,} 
{\it Living Rev. Relativity} {\bf 14} (2011) 8.

\bibitem{BFLW16} 
R. Bousso, Z. Fisher, S. Leichenauer, and A. C. Wall, 
{\it A Quantum Focussing Conjecture,}
{\it Phys. Rev. D} {\bf 93} (2016) 064044.

\bibitem{Miqueleto21}
J. L. Miqueleto and A. G. S. Landulfo,  
{\it Exact renormalization group, entanglement entropy, and black hole entropy,} 
{\it Phys. Rev. D} {\bf 103} (2021) 045012.

\bibitem{Zurek82} 
W. H. Zurek, 
{\it Entropy Evaporated by a Black Hole,}
{\it Phys. Rev. Lett.} {\bf 49} (1982) 1683.

\bibitem{Arderucio24}
B. Arderucio Costa,  
{\it Is the Page-time paradox paradoxical?} 
{\it Int. J. Mod. Phys. D} (2024) to appear
(https://arxiv.org/abs/2402.17815).

\end{thebibliography}
\end{document}